\documentclass[aps,prl,preprint,groupedaddress,twocolumn]{revtex4}
\usepackage{amsmath}
\usepackage{amssymb}
\usepackage{graphicx}
\usepackage{epsf}
\usepackage{mathtools}

\begin{document}

\title{Uncertainty, von Neumann Entropy, and Squeezing in a Bipartite State of Two-Level Atoms}

\author{Ram Narayan Deb\\
Email: debramnarayan1@gmail.com}
\affiliation{Bidhannagar College, EB-2, Sector-I, Salt Lake, Bidhannagar North, 24 Parganas(N), 700064, West Bengal, India}
\date{\today}

\begin{abstract}
We provide a generalized treatment of uncertainties, von Neumann entropy, and squeezing in entangled bipartite pure state of two-level atoms. We observe that when the bipartite state is entangled, though the von Neumann entropy of the composite state is less than those of the subsystems, as first recognized by Schr$\ddot{o}$dinger, but  the uncertainty of the composite state is greater than those of the subsystems  for certain ranges of the superposing constants of the quantum state. This is in contradiction with the prevailing idea  that the greater the entropy, the greater the uncertainty. Hence, for those ranges of the superposing constants of the quantum state, although the entropic inequalities are violated, the subsystems exhibit less disorder than the system as a whole. We also present generalized relationships between the von Neumann entropies of the subsystems and the uncertainty, spin squeezing, and spectroscopic squeezing parameters \textemdash used in the context of Ramsey spectroscopy \textemdash of the composite state. This provides a generalized operational measure of von Neumann entropy of entangled subsystems of two-level atoms. 
\\
{\it keywords: uncertainty, von Neumann entropy, two-level atom, spin squeezing, spectroscopic squeezing} 
\end{abstract}

\maketitle

\section{I INTRODUCTION}
Uncertainties in the simultaneous measurements of  incompatible observables of quantum systems, make quantum systems 
drastically different from classical ones. Uncertainty principle, first formulated by Heisenberg \cite{Heisenberg}, forms a central part of our understanding of quantum mechanics.
Many works have been done on further developments and applications of uncertainty principle, with the  advent and progress of quantum information theory \cite{Robertson}-\cite{Horodecki1}.

Von Neumann entropy, also called entanglement entropy, plays a crucial role in the quantification of quantum entanglement. Entropic manifestation of quantum entanglement forms a significant part of quantum information theory \cite{Horodecki1}. 

Spin squeezing is an important experimental tool that has potential applications in the detection of quantum entanglement, high-precision metrology, atomic clocks, quantum Fisher information, quantum chaos, quantum phase transitions, Bose-Einstein condensates, and  ion-trap experiments in quantum optics. Spin squeezing is the reduction of quantum fluctuations in collective spin observables of atomic systems, typically measured by the variance of the spin components. This is achieved by correlations between the spins, which can be generated through nonlinear interactions or entanglement \cite{Kitagawa, Jian}.

Spectroscopic squeezing, on the other hand, refers to the reduction of quantum fluctuations in the spectral response of an atomic ensemble. This is typically measured by the variance of the absorption or emission spectrum. It has significant applications in enhancing resolutions and precisions in optical measurements, which lead to more accurate determination of atomic and molecular energy levels. Squeezing of spectroscopic signals enables the detection of weaker signals  
and more precise characterization of atomic and molecular systems.  It enables the achievement of quantum-limited spectroscopy, where the measurement precision is limited only by the fundamental laws of quantum mechanics. It helps to develop more accurate atomic clocks, which rely on precise spectroscopic measurements. In quantum information processing, spectroscopic squeezing enables the development of more precise quantum gates and quantum computing architectures \cite{Wineland2, Jian}.

In this paper, we consider a general entangled state of two two-level atoms. We show that, although the von Neumann entropies of the entangled subsystems are always greater than that of the system as a whole \textemdash thus violating the entropic inequalities \cite{Horodecki1} \textemdash  the uncertainties of the subsystems are lower than those of the system as a whole for certain ranges of the superposition constants of the quantum state. This is in contradiction with the idea that, the more is the entropy, more is the uncertainty, and the more is the disorder. We also establish, analytically,  generalized relationships between the von Neumann entropies of the subsystems, and the uncertainties, spin squeezing, and spectroscopic squeezing parameters of the system as a whole. Now, since  spin squeezing and spectroscopic squeezing are experimentally measurable, this provides an operational measure of the entanglement entropy of the subsystems of two-level atoms. 

It is worthwhile to mention here that noise and decoherence are significant practical challenges in experiments involving spin and spectroscopic squeezing. Noise and decoherence can destroy spin coherence, introduce spin fluctuations, and  limit the squeezing factors in spin squeezing experiments. In spectroscopic squeezing experiments, they can broaden spectral lines, introduce spectral fluctuations, and limit squeezing factors. However, a lot of experimental and theoretical progress has been made in the study of spin and spectroscopic squeezing in many physical systems 
\textemdash for example, in ion traps and Bose \textemdash Einstein condensates [33 - 58].

In section II, we briefly discuss  two-level atoms, pseudo-spin operators, and spin and spectroscopic squeezing. In section III, we calculate the uncertainties of the pseudo-spin operators in two mutually orthogonal directions, in a plane perpendicular to the mean spin vector of the general entangled state of two two-level atoms, both for the composite state and the subsystems. We show that, although the von Neumann entropy of the composite state is less than those of the subsystems,  the uncertainties of the composite state are greater than those of the subsystems for certain ranges of the superposing constants of the quantum state. In section IV, we establish  general relationships between the von Neumann entropies of the subsystems and the uncertainties, spin squeezing, and spectroscopic squeezing parameters of the composite state. In section V, we present the conclusion.    

\subsection{II TWO-LEVEL ATOM, PSEUDO-SPIN OPERATORS, SPIN AND SPECTROSCOPIC SQUEEZING}
An atom has many electronic energy levels, but when it interacts with  external monochromatic electromagnetic radiation, it undergoes a transition from one energy level to another. In this case, we mainly focus on those two energy levels; hence, the atom is referred to as a two-level atom. 
Now, if among an assembly of $N$ such two-level atoms, the $n$-th atom has upper and lower energy levels denoted by 
$|u_n\rangle$ and $|l_n\rangle$, respectively, then we can construct the pseudo-spin operators (with $\hbar = 1$),
\begin{eqnarray}
\hat{J}_{n_x} &=& (1/2)\big(|u_n\rangle\langle l_n| + |l_n\rangle
\langle u_n|\big),\label{1.1a1}\nonumber\\
\hat{J}_{n_y} &=& (-i/2)\big(|u_n\rangle\langle l_n| - |l_n\rangle
\langle u_n|\big),\label{1.1a2}\nonumber\\
\hat{J}_{n_z} &=& 
(1/2)\big(|u_n\rangle\langle u_n| - |l_n\rangle\langle l_n|\big),\nonumber
\label{1.1}
\end{eqnarray}
such that,
\begin{equation}
 [\hat{J}_{n_x}, \hat{J}_{n_y}] = i\hat{J}_{n_z}, \nonumber
\label{1.2}
\end{equation}

 and two more relations with cyclic changes in $x$, $y$ and $z$ \cite{Itano}. Since these operators are not actual spin angular momentum operators, but satisfy the same commutation relations as those of spin-$\frac{1}{2}$ particles, they are referred to as pseudo-spin operators.
 For the entire system of two two-level atoms, we define the collective pseudo-spin operators
\begin{eqnarray}
\hat{J}_x = \sum_{i=1}^{2}\hat{J}_{i_{x}}, 
\hat{J}_y = \sum_{i=1}^{2}\hat{J}_{i_{y}}, ~~and~~
\hat{J}_z = \sum_{i=1}^{2}\hat{J}_{i_{z}},\nonumber
\label{1.3}
\end{eqnarray}
where each term in the summations is implicitly tensor-multiplied by identity operators acting on the other atom \cite {Itano}, \cite{Sakurai}.

The individual atomic operators satisfy
\begin{eqnarray}
\big[\hat{J}_{1_x}, \hat{J}_{2_y}\big] &=& 0,~
\big[\hat{J}_{1_x}, \hat{J}_{1_y}\big] = i\hat{J}_{1_z},\nonumber\\
\big[\hat{J}_{2_x}, \hat{J}_{2_y}\big] &=& i\hat{J}_{2_z},...,\nonumber
\label{1.4}
\end{eqnarray}
where the subscripts $1$ and $2$ are for atoms $1$ and $2$, respectively.

 As a consequence of these commutation
relations, the collective pseudo-spin operators 
$\hat{J}_x$, $\hat{J}_y$, and $\hat{J}_z$ satisfy
\begin{equation}
[\hat{J}_x , \hat{J}_y] = i\hat{J}_z,
\label{1.5}
\end{equation} 
and two more
relations with cyclic changes in $x$, $y$, and $z$.

The  eigenvectors of  $\hat{J}_{n_z}$ are denoted
as $|\frac{1}{2}\rangle$ and $|-\frac{1}{2}\rangle$, with eigenvalues $\frac{1}{2}$ and 
$-\frac{1}{2}$, respectively.

The use of these collective spin operators,  as discussed in Eq. (\ref{1.3}), is widely used in  problems of quantum optics \cite{Korbicz1}-\cite{Miyake2}.

In order to study the spin squeezing and spectroscopic squeezing properties of a quantum state, say 
$|\psi\rangle$, of a system of two-level atoms, we need to calculate the coordinate-frame independent inherent uncertainties in the components of $\hat{\mathbf{J}}$ along two mutually orthogonal directions in a plane perpendicular to the mean spin vector $\langle\hat{\mathbf{J}}\rangle$ of the system \cite{Kitagawa, Wineland2}, where
\begin{equation}
\langle\hat{\mathbf{J}}\rangle = \langle\hat{J}_{n_1}\rangle
\hat{n}_1 + \langle\hat{J}_{n_2}\rangle\hat{n}_2 + \langle\hat{J}_{n_3}\rangle\hat{n}_3.\nonumber
\label{3.8a2}
\end{equation}
$\hat{n}_1$, $\hat{n}_2$, and $\hat{n}_3$ are three mutually orthogonal unit vectors.
 The expectation values mentioned above are to be calculated with respect to the quantum state $|\psi\rangle$.
  
To understand this concretely, we take the directions of $\hat{n}_1$, $\hat{n}_2$, and $\hat{n}_3$ along the positive $x$, $y$, and $z$ axes, respectively, so that  
\begin{equation}
\langle\hat{\mathbf{J}}\rangle = \langle\hat{J}_x\rangle
\hat{i} + \langle\hat{J}_y\rangle\hat{j} + \langle\hat{J}_z\rangle\hat{k},\nonumber
\label{3.9}
\end{equation}
and hence the uncertainties are given by the standard deviations
\begin{equation}
\Delta J_{x,y} = \sqrt{\langle\psi|{\hat{J}_{x,y}}^2|\psi\rangle - 
\langle\psi|\hat{J}_{x,y}|\psi\rangle^2}. 
\label{3.10}
\end{equation}     

Now, since, for an arbitrary quantum state $|\psi\rangle$, the mean spin vector 
$\langle\hat{\mathbf{J}}\rangle$ points in an arbitrary direction in space, we conventionally rotate the coordinate system $\{x,y,z\}$ to $\{x^\prime, y^\prime, z^\prime\}$, such that 
$\langle\hat{\bf{J}}\rangle$ points along the $z^\prime$ axis, and calculate the standard deviations of $\hat{J}_{x^\prime}$ and $\hat{J}_{y^\prime}$ for the state $|\psi\rangle$. 
 Since the mean spin vector $\langle\hat{\mathbf{J}}\rangle$, after rotation of the coordinate frame, points along the $z^\prime$ axis, we have
\begin{eqnarray}
\langle\psi|\hat{J}_{x^\prime}|\psi\rangle = \langle\psi|\hat{J}_{y^\prime}|\psi\rangle = 0, \nonumber
\label{3.12}
\end{eqnarray}
and hence the standard deviations, as mentioned in Eq. (\ref{3.10}), reduce to
\begin{equation}
\Delta J_{x^\prime,y^\prime} = \sqrt{\langle\psi|{\hat{J}_{x^\prime,y^\prime}}^2|\psi\rangle}.
\label{3.13}
\end{equation}
In this paper, we calculate these coordinate-independent uncertainties for a general pure state of two two-level atoms, both for the composite state and for the individual entangled parts.

For a system of $N$ two-level atoms, the spin squeezing parameter, introduced by Kitagawa et al. \cite{Kitagawa, Jian}, is defined as
\begin{eqnarray}
\xi_s = \sqrt{\frac{2}{j}}\Delta J_{\overrightarrow{n}_{\perp}}, \nonumber
\label{3.56}
\end{eqnarray}
 where $\overrightarrow{n}_{\perp}$ lies in the plane perpendicular to the mean spin vector 
 $\langle\hat{\mathbf{J}}\rangle$, and $j = \frac{N}{2}$. Now, conventionally, we take $\overrightarrow{n}_{\perp}$ as the directions of $x^\prime$ and $y^\prime$ axes, and hence, the spin squeezing parameters become
\begin{eqnarray}
\xi_{s_x} = \sqrt{\frac{2}{j}}\Delta J_{x^\prime}~~and~~\xi_{s_y} = \sqrt{\frac{2}{j}}\Delta J_{y^\prime}. 
\label{3.56a1}
\end{eqnarray}
If, for a quantum state, $\xi_{s_x} < 1$, then we have $\xi_{s_y} > 1$, respecting Heisenberg's uncertainty principle, and we conclude that we have spin squeezing in the $x$-quadrature \textemdash and vice versa. The degree of spin squeezing is quantified by this spin squeezing parameter. 
The smaller $\xi_s$ is (i.e., the more  $\xi_s < 1$), the greater the spin squeezing.

The spectroscopic squeezing parameter $\xi_R$, introduced by Wineland et al. \cite{Wineland2, Jian}, is given by
\begin{eqnarray}
\xi_R = \sqrt{N}\frac{\Delta J_{\overrightarrow{n}_{\perp}}}{|\langle\hat{\mathbf{J}}\rangle|}. 
\label{3.58}\nonumber
\end{eqnarray}
In a similar fashion, as mentioned above, we take the directions of $\overrightarrow{n}_{\perp}$ as the directions of $x^\prime$ and $y^\prime$ axes, and hence, the spectroscopic squeezing parameters become
\begin{eqnarray}
\xi_{R_x} = \sqrt{N}\frac{\Delta J_{x^\prime}}{|\langle\hat{\mathbf{J}}\rangle|}~~and~~\xi_{R_y} = \sqrt{N}\frac{\Delta J_{y^\prime}}{|\langle\hat{\mathbf{J}}\rangle|}.
\label{3.58a1}
\end{eqnarray}
If $\xi_{R_x}$ or $\xi_{R_y}$ $< 1$, we have spectroscopic squeezing in the system. Spectroscopic squeezing is quantified using these parameters. The smaller the value of $\xi_R < 1$, the greater the spectroscopic squeezing.

We now proceed to study the coordinate-independent inherent uncertainties and the von Neumann entropy 
of a general system of two two-level atoms.

\subsection{III UNCERTAINTIES AND VON NEUMANN ENTROPY IN A SYSTEM OF TWO TWO-LEVEL ATOMS}

 A general pure state of two systems, say $A$ and $B$, is given by 
\begin{eqnarray}
|\psi_{AB}\rangle = \sum_{i=1}^{2} c_i |e^i_{A}\rangle \otimes |e^i_{B}\rangle,\nonumber
\label{3.1}
\end{eqnarray}
where $|\psi_{AB}\rangle$ $\in$ $H_A \otimes H_B$, $c_i$ are the Schmidt coefficients, and 
$\{|e^i_{A}\rangle \otimes |e^i_{B}\rangle\}$ form an orthonormal product basis. 
The Schmidt coefficients $c_i$ satisfy 
\begin{equation}
\sum_{i=1}^{2} {c_i}^2= 1.
\label{3.2} 
\end{equation}

We now consider a system of two entangled two-level atoms, denoted as atoms $a$ and $b$. To construct the composite quantum pure state $|\psi_{ab}\rangle$ of the two atoms, we take the orthonormal product basis, composed of generalized superpositions of $|\frac{1}{2}\rangle$ and $|-\frac{1}{2}\rangle$, given by 
\begin{eqnarray}
\Big{\{}\big{(}c_3|\frac{1}{2}\rangle + c_4|-\frac{1}{2}\rangle\big{)} \otimes \big{(}c_5|\frac{1}{2}\rangle + c_6|-\frac{1}{2}\rangle\big{)}\Big{\}}\nonumber
\end{eqnarray}
and 
\begin{eqnarray}
\Big{\{}\big{(}c_4|\frac{1}{2}\rangle - c_3|-\frac{1}{2}\rangle\big{)} \otimes \big{(}c_6|\frac{1}{2}\rangle - c_5|-\frac{1}{2}\rangle\big{)}\Big{\}},\nonumber
\end{eqnarray}
where $|\frac{1}{2}\rangle$ and $|-\frac{1}{2}\rangle$ are the eigenvectors of $\hat{J}_{a_z}$ and $\hat{J}_{b_z}$, corresponding to eigenvalues $\frac{1}{2}$ and $-\frac{1}{2}$,
respectively.  For the sake of simplicity, $c_3$, $c_4$, $c_5$, and $c_6$ are assumed to be real.
Since the basis vectors are normalized, we have
\begin{eqnarray}
{c_3}^2 + {c_4}^2 = 1 \nonumber\label{3.2a1}
\end{eqnarray}
and 
\begin{eqnarray}
{c_5}^2 + {c_6}^2 = 1 \nonumber\label{3.2a2}.
\end{eqnarray}

 Therefore, the complete form of $|\psi_{ab}\rangle$ is
\begin{eqnarray}
&&|\psi_{ab}\rangle = c_1\bigg{[}\big{(}c_3\Big{|}\frac{1}{2}\Big{\rangle} + c_4\Big{|}-\frac{1}{2}\Big{\rangle}\big{)} \otimes\nonumber\\ &&\big{(}c_5\Big{|}\frac{1}{2}\Big{\rangle} + c_6\Big{|}-\frac{1}{2}\Big{\rangle}\big{)} \bigg{]} + \nonumber\\ &&c_2\bigg{[} \big{(}c_4\Big{|}\frac{1}{2}\Big{\rangle} - c_3\Big{|}-\frac{1}{2}\Big{\rangle}\big{)}\nonumber\\
 &&\otimes \big{(}c_6\Big{|}\frac{1}{2}\Big{\rangle} - c_5\Big{|}-\frac{1}{2}\Big{\rangle}\big{)} \bigg{]}.\nonumber
 \label{3.3} 
\end{eqnarray}

The above quantum state can be written as
\begin{eqnarray}
&&|\psi_{ab}\rangle = \Big{(}c_1c_3c_5 + c_2c_4c_6\Big{)}\Big{|}\frac{1}{2}\Big{\rangle}\otimes \Big{|}\frac{1}{2}\Big{\rangle} + \nonumber\\
&&\Big{(}c_1c_3c_6 - c_2c_4c_5\Big{)} \Big{|}\frac{1}{2}\Big{\rangle}\otimes \Big{|}-\frac{1}{2}\Big{\rangle} + \nonumber\\ 
&&\Big{(}c_1c_4c_5 - c_2c_3c_6\Big{)} \Big{|}-\frac{1}{2}\Big{\rangle}\otimes \Big{|}\frac{1}{2}\Big{\rangle}
+ \nonumber\\ &&\Big{(}c_1c_4c_6 + c_2c_3c_5\Big{)} \Big{|}-\frac{1}{2}\Big{\rangle}\otimes \Big{|}-\frac{1}{2}\Big{\rangle}.
\label{3.4}
\end{eqnarray}

To simplify the calculations, we let
\begin{eqnarray}
\Big{(}c_1c_3c_5 + c_2c_4c_6\Big{)} &=& A,\label{3.5}\\
\Big{(}c_1c_3c_6 - c_2c_4c_5\Big{)} &=& B, \label{3.6}\\
\Big{(}c_1c_4c_5 - c_2c_3c_6\Big{)} &=& C, \label{3.7}\\
\Big{(}c_1c_4c_6 + c_2c_3c_5\Big{)} &=& D, \label{3.8}
\end{eqnarray}
where
\begin{eqnarray}
A^2 + B^2 +C^2 + D^2 = 1.
\label{3.8a1}
\end{eqnarray}
Therefore, Eq. (\ref{3.4}) becomes
\begin{eqnarray}
&&|\psi_{ab}\rangle = A \Big{|}\frac{1}{2}\Big{\rangle}\otimes \Big{|}\frac{1}{2}\Big{\rangle} + 
B \Big{|}\frac{1}{2}\Big{\rangle}\otimes \Big{|}-\frac{1}{2}\Big{\rangle} + \nonumber\\ 
&& C \Big{|}-\frac{1}{2}\Big{\rangle}\otimes \Big{|}\frac{1}{2}\Big{\rangle}
+ D \Big{|}-\frac{1}{2}\Big{\rangle}\otimes \Big{|}-\frac{1}{2}\Big{\rangle}.
\label{3.4a11}
\end{eqnarray}

We now proceed to calculate the coordinate independent inherent uncertainties, 
$\Delta J_{x^\prime,y^\prime}$, as mentioned in the previous section. To that end, we first calculate 
$\langle\psi_{ab}|\hat{J}_{x}|\psi_{ab}\rangle$, $\langle\psi_{ab}|\hat{J}_{y}|\psi_{ab}\rangle$, 
and $\langle\psi_{ab}|\hat{J}_{z}|\psi_{ab}\rangle$. We find that
\begin{eqnarray}
\langle\psi_{ab}|\hat{J}_{x}|\psi_{ab}\rangle &=& (A+D)(B+C),\label{3.14}\\ 
\langle\psi_{ab}|\hat{J}_{y}|\psi_{ab}\rangle &=& 0, \label{3.15} \\
\langle\psi_{ab}|\hat{J}_{z}|\psi_{ab}\rangle &=& A^2 - D^2,
\label{3.16}
\end{eqnarray}
and hence, the square of the magnitude of the mean spin vector is
\begin{eqnarray}
&&|\langle\hat{\mathbf{J}}\rangle|^2 = \langle\psi_{ab}|\hat{J}_{x}|\psi_{ab}\rangle^2 + 
\langle\psi_{ab}|\hat{J}_{z}|\psi_{ab}\rangle^2\nonumber\\
&&= A^2(A^2 + B^2 + C^2 + 2BC) + \nonumber\\
&& D^2(B^2 + C^2 + D^2 - 2A^2 + 2BC) + \nonumber\\
&& 2AD(B + C)^2.
\label{3.16a1}
\end{eqnarray} 

We notice from Eqs. (\ref{3.14}) to (\ref{3.16}) that the mean spin vector 
$\langle\hat{\mathbf{J}}\rangle$ lies in the $x-z$ plane. To align it along the $z^\prime$ direction, we perform the following rotation:
\begin{eqnarray}
\hat{J}_{x^\prime} &=& \hat{J}_x\cos\theta - \hat{J}_z\sin\theta\label{3.17}\\
\hat{J}_{y^\prime} &=& \hat{J}_y\label{3.18}\\
\hat{J}_{z^\prime} &=& \hat{J}_x\sin\theta + \hat{J}_z\cos\theta,\nonumber\label{3.19}
\end{eqnarray}
where
\begin{eqnarray}
\cos\theta &=& \frac{\langle\hat{J}_z\rangle}
{|\langle\hat{\mathbf{J}}\rangle|}.
\label{3.20}
\end{eqnarray}
It can be verified using Eqs. (\ref{3.15}), (\ref{3.17}), (\ref{3.18}), and (\ref{3.20}) that
\begin{eqnarray}
\langle\psi_{ab}|\hat{J}_{x^\prime}|\psi_{ab}\rangle &=& 0, \nonumber\label{3.20a}\\ 
\langle\psi_{ab}|\hat{J}_{y^\prime}|\psi_{ab}\rangle &=& 0, \nonumber\label{3.20b} 
\end{eqnarray}
and hence, the mean spin vector $\langle\hat{\mathbf{J}}\rangle$ is now aligned along the $z^\prime$-axis.

Using Eqs. (\ref{3.17}) and (\ref{3.20}), we obtain
\begin{eqnarray}
&&\langle\hat{J}_{x^\prime}^2\rangle = \frac{1}{|\langle\hat{\mathbf{J}}\rangle|^2}\Big{[} \langle \hat{J}_x^2\rangle\langle\hat{J}_z\rangle^2 + \langle \hat{J}_z^2\rangle\langle\hat{J}_x\rangle^2\nonumber\\
&& - \langle\hat{J}_x\hat{J}_z + \hat{J}_z\hat{J}_x\rangle\langle\hat{J}_x\rangle\langle\hat{J}_z\rangle\Big{]}. 
\label{3.21}
\end{eqnarray}
We now calculate $\langle\psi_{ab}|\hat{J}_x^2|\psi_{ab}\rangle$, $\langle\psi_{ab}|\hat{J}_z^2|\psi_{ab}\rangle$, and $\langle\psi_{ab}|\hat{J}_x\hat{J}_z + \hat{J}_z\hat{J}_x|\psi_{ab}\rangle$ for the quantum state in Eq. (\ref{3.3}), using the expression for $|\psi_{ab}\rangle$ given in 
Eq. (\ref{3.4a11}).

Now,
\begin{eqnarray}
\hat{J}_x = \frac{1}{2}\Big{(} \hat{J}_{+} + \hat{J}_{-} \Big{)}.\nonumber
\label{3.4a12}
\end{eqnarray}
Therefore,
\begin{eqnarray}
\hat{J}_x^2 = \frac{1}{4}\Big{(} \hat{J}_{+}^2 + \hat{J}_{-}^2 + \hat{{J}_{+}}\hat{{J}_{-}} + \hat{{J}_{-}}\hat{{J}_{+}} \Big{)},
\label{3.4a13}
\end{eqnarray}
and
\begin{eqnarray}
&&\hat{J}_x\hat{J}_z + \hat{J}_z\hat{J}_x = \frac{1}{2}\Big{(}\hat{J}_+\hat{J}_z + 
\hat{J}_z\hat{J}_+ + \nonumber\\ 
&& \hat{J}_-\hat{J}_z + \hat{J}_z\hat{J}_-\Big{)}.
\label{3.4a13b}
\end{eqnarray}
Since we are dealing here only with two atoms, $a$ and $b$, we have
\begin{eqnarray}
\hat{J}_{+} &=& \hat{J}_{a_+} + \hat{J}_{b_+}, \nonumber\label{3.4a14}\\
\hat{J}_{-} &=& \hat{J}_{a_-} + \hat{J}_{b_-}, \nonumber\label{3.4a15}\\
\hat{J}_{z} &=& \hat{J}_{a_z} + \hat{J}_{b_z}. \label{3.4a15a1}
\end{eqnarray}
Therefore,
\begin{eqnarray}
&&\hat{J}_{+}^2 = \hat{J}_{a_+}^2 + \hat{J}_{b_+}^2 + 2\hat{J}_{a_+}\hat{J}_{b_+},\label{3.4a16}\\ 
&&\hat{J}_{+}\hat{J}_{-} + \hat{J}_{-}\hat{J}_{+} = \hat{J}_{a_+}\hat{J}_{a_-} + \hat{J}_{a_-}
\hat{J}_{a_+}\nonumber\\
&& + \hat{J}_{b_+}\hat{J}_{b_-} + \hat{J}_{b_-}\hat{J}_{b_+} + 2\hat{J}_{a_+}\hat{J}_{b_-} + 
2\hat{J}_{a_-}\hat{J}_{b_+},\nonumber\\
\label{3.4a15a2}
\end{eqnarray}
and
\begin{eqnarray}
&&\hat{J}_+\hat{J}_z + \hat{J}_z\hat{J}_+ = \hat{J}_{a_+}\hat{J}_{a_z} + \hat{J}_{a_z}\hat{J}_{a_+}
+ \nonumber\\
&&\hat{J}_{b_+}\hat{J}_{b_z} + \hat{J}_{b_z}\hat{J}_{b_+} + 2\Big{(}\hat{J}_{a_+}\hat{J}_{b_z} + \hat{J}_{a_z}\hat{J}_{b_+}\Big{)}.\nonumber\\
\label{3.4a15a3}
\end{eqnarray}
Now, as $A$, $B$, $C$, and $D$ are real, we have, using Eqs. (\ref{3.4a11}) and (\ref{3.4a16}),
\begin{eqnarray}
&&\langle\psi_{ab}|\hat{J}_{+}^2|\psi_{ab}\rangle =
\bigg{[}\Big{\langle}\frac{1}{2}\Big{|}\otimes \Big{\langle}\frac{1}{2}\Big{|} A +\nonumber\\ 
&&\Big{\langle}-\frac{1}{2}\Big{|}\otimes \Big{\langle}\frac{1}{2}\Big{|} B + 
\Big{\langle}\frac{1}{2}\Big{|}\otimes \Big{\langle}-\frac{1}{2}\Big{|} C + \nonumber\\
&&\Big{\langle}-\frac{1}{2}\Big{|}\otimes \Big{\langle}-\frac{1}{2}\Big{|} D \bigg{]}\bigg{|}\Big{(}\hat{J}_{a_+}^2 + \hat{J}_{b_+}^2 + \nonumber\\
&& 2\hat{J}_{a_+}\hat{J}_{b_+}\Big{)}\bigg{|}
\bigg{[} A \Big{|}\frac{1}{2}\Big{\rangle}\otimes \Big{|}\frac{1}{2}\Big{\rangle} +\nonumber\\ 
&& B \Big{|}\frac{1}{2}\Big{\rangle}\otimes \Big{|}-\frac{1}{2}\Big{\rangle} +  
 C \Big{|}-\frac{1}{2}\Big{\rangle}\otimes \Big{|}\frac{1}{2}\Big{\rangle} \nonumber\\
&& + D \Big{|}-\frac{1}{2}\Big{\rangle}\otimes \Big{|}-\frac{1}{2}\Big{\rangle}\bigg{]}\nonumber\\
&& = 2 A D. \label{3.4a17}
\end{eqnarray}
So,
\begin{eqnarray}
&&\langle\psi_{ab}|\hat{J}_{-}^2|\psi_{ab}\rangle = \langle\psi_{ab}|\hat{J}_{+}^2|\psi_{ab}\rangle^*\nonumber\\
&& = 2 A D.
\label{3.4a18}
\end{eqnarray}
Using Eqs. (\ref{3.4a11}) and (\ref{3.4a15a2}), we obtain
\begin{eqnarray}
&&\langle\psi_{ab}|\Big{(}\hat{J}_{+}\hat{J}_{-} + \hat{J}_{-}\hat{J}_{+}\Big{)}|\psi_{ab}\rangle = \nonumber\\
&&\bigg{[}\Big{\langle}\frac{1}{2}\Big{|}\otimes \Big{\langle}\frac{1}{2}\Big{|} A + 
\Big{\langle}-\frac{1}{2}\Big{|}\otimes \Big{\langle}\frac{1}{2}\Big{|} B + \nonumber\\
&&\Big{\langle}\frac{1}{2}\Big{|}\otimes \Big{\langle}-\frac{1}{2}\Big{|} C + 
\Big{\langle}-\frac{1}{2}\Big{|}\otimes \Big{\langle}-\frac{1}{2}\Big{|} D \bigg{]}\bigg{|}
\nonumber\\
&&\Big{(}\hat{J}_{a_+}\hat{J}_{a_-} + \hat{J}_{a_-}\hat{J}_{a_+}
 + \hat{J}_{b_+}\hat{J}_{b_-} + \hat{J}_{b_-}\hat{J}_{b_+} + \nonumber\\
 && 2\hat{J}_{a_+}\hat{J}_{b_-} + 2\hat{J}_{a_-}\hat{J}_{b_+}\Big{)}\bigg{|}
\bigg{[} A \Big{|}\frac{1}{2}\Big{\rangle}\otimes \Big{|}\frac{1}{2}\Big{\rangle} +\nonumber\\ 
&& B \Big{|}\frac{1}{2}\Big{\rangle}\otimes \Big{|}-\frac{1}{2}\Big{\rangle} +  
 C \Big{|}-\frac{1}{2}\Big{\rangle}\otimes \Big{|}\frac{1}{2}\Big{\rangle} \nonumber\\
&& + D \Big{|}-\frac{1}{2}\Big{\rangle}\otimes \Big{|}-\frac{1}{2}\Big{\rangle}\bigg{]}\nonumber\\
&& = 2 \Big{(} A^2 + B^2 + C^2 + D^2 + 2BC\Big{)}. \label{3.4a19}
\end{eqnarray}
Now, by virtue of Eq. (\ref{3.8a1}), Eq. (\ref{3.4a19}) reduces to
\begin{eqnarray}
\langle\psi_{ab}|\Big{(}\hat{J}_{+}\hat{J}_{-} + \hat{J}_{-}\hat{J}_{+}\Big{)}|\psi_{ab}\rangle = 2\Big{(}
1 + 2BC \Big{)}.\nonumber\\
\label{3.4a20}
\end{eqnarray}
Therefore, using Eqs. (\ref{3.4a13}), (\ref{3.4a17}), (\ref{3.4a18}), and (\ref{3.4a20}), we get
\begin{eqnarray}
\langle\psi_{ab}|\hat{J}_x^2|\psi_{ab}\rangle = \frac{1}{2}\Big{(}1 + 2AD + 2BC\Big{)}. \label{3.22}
\end{eqnarray}
Now, using Eqs. (\ref{3.4a11}) and (\ref{3.4a15a1}), we obtain
\begin{eqnarray}
&&\langle\psi_{ab}|\hat{J}_{z}^2|\psi_{ab}\rangle =
\bigg{[}\Big{\langle}\frac{1}{2}\Big{|}\otimes \Big{\langle}\frac{1}{2}\Big{|} A +\nonumber\\ 
&&\Big{\langle}-\frac{1}{2}\Big{|}\otimes \Big{\langle}\frac{1}{2}\Big{|} B + 
\Big{\langle}\frac{1}{2}\Big{|}\otimes \Big{\langle}-\frac{1}{2}\Big{|} C + \nonumber\\
&&\Big{\langle}-\frac{1}{2}\Big{|}\otimes \Big{\langle}-\frac{1}{2}\Big{|} D \bigg{]}\bigg{|}\Big{(}\hat{J}_{a_z}^2 + \hat{J}_{b_z}^2 + \nonumber\\
&& 2\hat{J}_{a_z}\hat{J}_{b_z}\Big{)}\bigg{|}
\bigg{[} A \Big{|}\frac{1}{2}\Big{\rangle}\otimes \Big{|}\frac{1}{2}\Big{\rangle} +\nonumber\\ 
&& B \Big{|}\frac{1}{2}\Big{\rangle}\otimes \Big{|}-\frac{1}{2}\Big{\rangle} +  
 C \Big{|}-\frac{1}{2}\Big{\rangle}\otimes \Big{|}\frac{1}{2}\Big{\rangle} \nonumber\\
&& + D \Big{|}-\frac{1}{2}\Big{\rangle}\otimes \Big{|}-\frac{1}{2}\Big{\rangle}\bigg{]}\nonumber\\
&& = A^2 + D^2. \label{3.23}
\end{eqnarray}
Similarly, using Eqs. (\ref{3.4a11}) and (\ref{3.4a15a3}), we get
\begin{eqnarray}
&&\langle\psi_{ab}|\Big{(}\hat{J}_+\hat{J}_z + \hat{J}_z\hat{J}_+\Big{)}|\psi_{ab}\rangle = \big{(}A - 
D\big{)}\times\nonumber\\
&&\big{(}B + C\big{)}.
\label{3.23a1}
\end{eqnarray}
Therefore, 
\begin{eqnarray}
&&\langle\psi_{ab}|\Big{(}\hat{J}_-\hat{J}_z + \hat{J}_z\hat{J}_-\Big{)}|\psi_{ab}\rangle = 
\nonumber\\
&&\langle\psi_{ab}|\Big{(}\hat{J}_+\hat{J}_z + \hat{J}_z\hat{J}_+\Big{)}|\psi_{ab}\rangle^* = \nonumber\\
&&\big{(}A - D\big{)}\big{(}B + C\big{)}.
\label{3.23a2}
\end{eqnarray}
So, using Eqs. (\ref{3.4a13b}), (\ref{3.23a1}), and (\ref{3.23a2}), we obtain
\begin{eqnarray}
&&\langle\psi_{ab}|\Big{(}\hat{J}_x\hat{J}_z + \hat{J}_z\hat{J}_x\Big{)}|\psi_{ab}\rangle = 
\nonumber\\
&&(A-D)(B+C).
\label{3.24}
\end{eqnarray}
Therefore, using the expressions for $\langle\hat{J}_x\rangle$, $\langle\hat{J}_z\rangle$,
$|\langle\hat{\mathbf{J}}\rangle|^2$, $\langle\hat{J}_x^2\rangle$, $\langle\hat{J}_z^2\rangle$,  and 
$\langle\hat{J}_x\hat{J}_z + \hat{J}_z\hat{J}_x\rangle$ from Eqs. (\ref{3.14}), (\ref{3.16}), (\ref{3.16a1}), (\ref{3.22}),  (\ref{3.23}),  and (\ref{3.24}), respectively, in Eq. (\ref{3.21}), and simplifying,  we get
\begin{eqnarray}
\langle\psi_{ab}|\hat{J}_{x^\prime}^2|\psi_{ab}\rangle = \frac{1}{2}\big{(}A + D\big{)}^2.
\label{3.25}
\end{eqnarray}
So, using Eqs. (\ref{3.13}) and (\ref{3.25}), we have
\begin{eqnarray}
\Delta{J}_{x^\prime}^2 = \frac{1}{2}\big{(}A + D\big{)}^2.
\label{3.26}
\end{eqnarray}
Now,
\begin{eqnarray}
\hat{J}_y = \frac{1}{2i}\Big{(} \hat{J}_{+} - \hat{J}_{-} \Big{)}.\nonumber
\label{3.26b1}
\end{eqnarray}
Therefore,
\begin{eqnarray}
&&\langle\psi_{ab}|\hat{J}_y^2|\psi_{ab}\rangle = \frac{1}{4}\langle\psi_{ab}|\Big{(} \hat{{J}_{+}}\hat{{J}_{-}} + \hat{{J}_{-}}\hat{{J}_{+}} \nonumber\\
&&- \hat{J}_{+}^2 - \hat{J}_{-}^2 \Big{)}|\psi_{ab}\rangle.
\label{3.26b2}
\end{eqnarray}
So, using Eqs. (\ref{3.4a17}), (\ref{3.4a18}), and (\ref{3.4a20}) in Eq. (\ref{3.26b2}), and simplifying, we obtain 
\begin{eqnarray}
\langle\psi_{ab}|\hat{J}_{y}^2|\psi_{ab}\rangle = \frac{1}{2}\Big{(} 1 - 2AD + 2BC\Big{)}.
\label{3.27}
\end{eqnarray}
Therefore, using Eqs. (\ref{3.13}), (\ref{3.18}), and (\ref{3.27}), we get
\begin{eqnarray}
\Delta{J}_{y^\prime}^2 &=& \langle\psi_{ab}|\hat{J}_{y^\prime}^2|\psi_{ab}\rangle \nonumber\\
&=& \frac{1}{2}\Big{(} 1 - 2AD + 2BC\Big{)}.
\label{3.28}
\end{eqnarray}
Now, using Eqs. (\ref{3.5}), (\ref{3.6}), (\ref{3.7}), (\ref{3.8}), (\ref{3.26}), and (\ref{3.28}), we get
\begin{eqnarray}
&&\Delta{J}_{x^\prime} = \sqrt{\Big{(} \frac{1}{2} + c_1c_2 \Big{)}} \Big{(} c_3c_5 + c_4c_6\Big{)},
\label{3.29}\\
&&\Delta{J}_{y^\prime} = \sqrt{\Big{(} \frac{1}{2} - c_1c_2 \Big{)}}.
\label{3.30}
\end{eqnarray}
These are the coordinate frame-independent inherent uncertainties of the composite state 
$|\psi_{ab}\rangle$ of the system.

We now proceed to calculate similar uncertainties in the atomic operators
$\hat{J}_{a_{x^\prime}}$, $\hat{J}_{a_{y^\prime}}$, $\hat{J}_{b_{x^\prime}}$, and 
$\hat{J}_{b_{y^\prime}}$ of atoms $a$ and $b$, respectively.

Now, the reduced density matrix $\hat{\rho}_a$ for atom $a$ is
\begin{eqnarray}
&&\hat{\rho}_a = c_1^2 \Big{[} c_3\Big{|}\frac{1}{2}\Big{\rangle} + c_4\Big{|}-\frac{1}{2}\Big{\rangle} \Big{]} \Big{[}\Big{\langle}\frac{1}{2}\Big{|}c_3 + \nonumber\\
&& \Big{\langle}-\frac{1}{2}\Big{|}c_4\Big{]} + c_2^2 \Big{[} c_4\Big{|}\frac{1}{2}\Big{\rangle} - c_3\Big{|}-\frac{1}{2}\Big{\rangle} \Big{]} \nonumber\\
&& \Big{[}\Big{\langle}\frac{1}{2}\Big{|}c_4 - \Big{\langle}-\frac{1}{2}\Big{|}c_3\Big{]}.\nonumber
\label{3.30a1} 
\end{eqnarray}
We find that
\begin{eqnarray}
&&\langle\hat{J}_{a_x}\rangle = tr(\hat{\rho}_a\hat{J}_{a_x}) = c_3c_4 (c_1^2 -c_2^2),\label{3.31}\\
&&\langle\hat{J}_{a_y}\rangle = tr(\hat{\rho}_a\hat{J}_{a_y}) = 0, \label{3.32}\\
&&\langle\hat{J}_{a_z}\rangle = tr(\hat{\rho}_a\hat{J}_{a_z}) = \frac{1}{2}(c_1^2 - c_2^2)(c_3^2-c_4^2).\label{3.33}\nonumber\\  
\end{eqnarray}
So, the mean spin vector $\langle\hat{\mathbf{J}}_a\rangle$ lies in the $x-z$ plane. Therefore, we perform the rotation
\begin{eqnarray}
\hat{J}_{a_{x^\prime}} &=& \hat{J}_{a_x}\cos\theta_1 - \hat{J}_{a_z}\sin\theta_1,\label{3.34}\\
\hat{J}_{a_{y^\prime}} &=& \hat{J}_{a_y},\label{3.35}\\
\hat{J}_{a_{z^\prime}} &=& \hat{J}_{a_x}\sin\theta_1 + \hat{J}_{a_z}\cos\theta_1,\nonumber\label{3.36}
\end{eqnarray}
where
\begin{eqnarray}
\cos\theta_1 &=& \frac{\langle\hat{J}_{a_z}\rangle}
{|\langle\hat{\mathbf{J}}_a\rangle|}.
\label{3.37}
\end{eqnarray}
It can be verified, using Eqs. (\ref{3.32}), (\ref{3.34}), (\ref{3.35}), and (\ref{3.37}), that
\begin{eqnarray}
\langle\hat{J}_{a_{x^\prime}}\rangle &=& 0, \nonumber\label{3.38}\\ 
\langle\hat{J}_{a_{y^\prime}}\rangle &=& 0, \nonumber\label{3.39} 
\end{eqnarray}
and hence, the mean spin vector $\langle\hat{\mathbf{J}}_a\rangle$ is now along the $z^\prime$-axis.
Therefore,
\begin{equation}
\Delta J_{a_{x^\prime,y^\prime}} = \sqrt{\langle{\hat{J}_{a_{x^\prime,y^\prime}}}^2\rangle}.
\label{3.40}
\end{equation}

Similar to Eq. (\ref{3.21}), we have, using Eqs. (\ref{3.34}) and (\ref{3.37}),
\begin{eqnarray}
&&\langle\hat{J}_{a_{x^\prime}}^2\rangle = \frac{1}{|\langle\hat{\mathbf{J}}_a\rangle|^2}\Big{[} \langle \hat{J}_{a_x}^2\rangle\langle\hat{J}_{a_z}\rangle^2 + \langle \hat{J}_{a_z}^2\rangle\langle\hat{J}_{a_x}\rangle^2\nonumber\\
&& - \langle\hat{J}_{a_x}\hat{J}_{a_z} + \hat{J}_{a_z}\hat{J}_{a_x}\rangle\langle\hat{J}_{a_x}\rangle\langle\hat{J}_{a_z}\rangle\Big{]}. 
\label{3.41}
\end{eqnarray}
We find that
\begin{eqnarray}
&&\langle\hat{J}_{a_x}^2\rangle = tr(\hat{\rho}_a\hat{J}_{a_x}^2) = \frac{1}{4},\label{3.42}\\
&&\langle\hat{J}_{a_y}^2\rangle = tr(\hat{\rho}_a\hat{J}_{a_y}^2) = \frac{1}{4},\label{3.43}\\
&&\langle\hat{J}_{a_z}^2\rangle = tr(\hat{\rho}_a\hat{J}_{a_z}^2) = \frac{1}{4},\label{3.44}\\
&&\langle\hat{J}_{a_x}\hat{J}_{a_z} + \hat{J}_{a_z}\hat{J}_{a_x}\rangle =\nonumber\\
&&tr\{\hat{\rho}_a (\hat{J}_{a_x}\hat{J}_{a_z} + \hat{J}_{a_z}\hat{J}_{a_x})\} = 0.\label{3.45}
\end{eqnarray}
Therefore, using Eqs. (\ref{3.31}) to (\ref{3.33}) and (\ref{3.40}) to (\ref{3.45}), we get
\begin{eqnarray}
\Delta J_{a_{x^\prime}} &=& \frac{1}{2},\label{3.46}\\
\Delta J_{a_{y^\prime}} &=& \frac{1}{2}.\label{3.47}
\end{eqnarray}
Similarly, it can be shown that, for atom $b$, the uncertainties in $\hat{J}_{b_{x^\prime}}$ and 
$\hat{J}_{b_{y^\prime}}$ are
\begin{eqnarray}
\Delta J_{b_{x^\prime}} &=& \frac{1}{2},\label{3.48}\\
\Delta J_{b_{y^\prime}} &=& \frac{1}{2}.\label{3.49}
\end{eqnarray}
Now, the von Neumann entropy of the composite state in Eq. (\ref{3.3}) is zero, as it is a pure state.
Therefore,
\begin{eqnarray}
S_{a,b} = 0.\nonumber
\label{3.49a1}
\end{eqnarray} 
But the von Neumann entropies of atoms $a$ and $b$, which are entangled, are
\begin{eqnarray}
S_{a} = S_{b} = - c_1^2 log_2c_1^2 - c_2^2 log_2c_2^2.
\label{3.50}
\end{eqnarray}

We now plot the uncertainties in Eqs. (\ref{3.29}) and (\ref{3.30}), and the von Neumann entropy in 
Eq. (\ref{3.50}), with respect to the constant $c_1$ in FIG. 1. 
\begin{figure}
\begin{center}
\hspace*{-1.0cm}
\includegraphics[width=9.5cm]{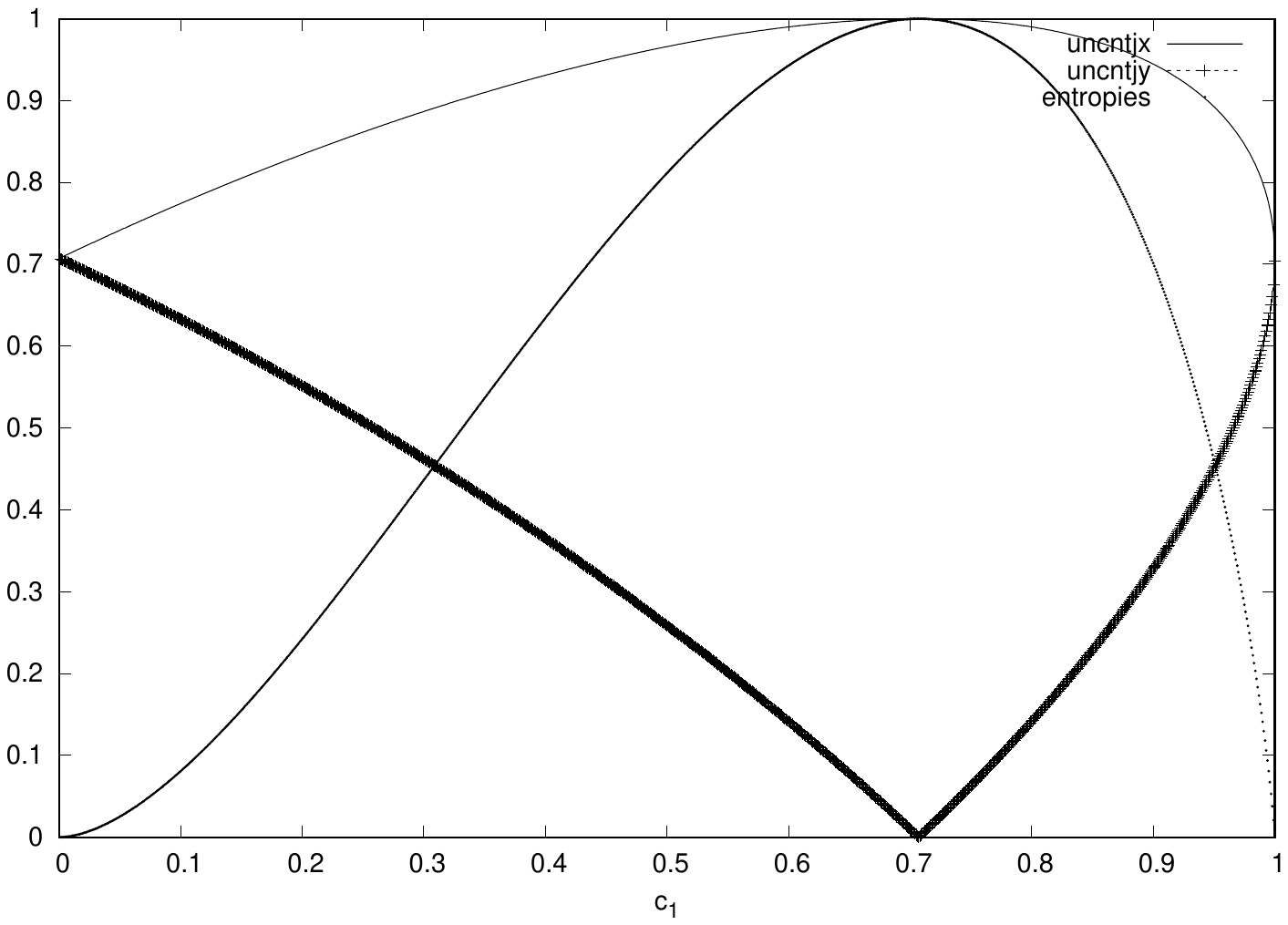}
\caption{Plot of the uncertainties  $\Delta J_{x^\prime}$, $\Delta J_{y^\prime}$, and the von Neumann entropies $S_{a}$ and $S_{b}$ of atoms $a$ and $b$ with respect to the constant $c_1$. Here, we have taken $c_3 = c_4 = c_5 = c_6 = \frac{1}{\sqrt{2}}$. The constant $c_1$ is plotted along the horizontal axis, while $\Delta J_{x^\prime}$, $\Delta J_{y^\prime}$, and the entropies $S_{a}$ and  $S_{b}$ are plotted along the vertical axis. The thin line, thick line, and dotted line represent 
$\Delta J_{x^\prime}$, $\Delta J_{y^\prime}$, and $S_{a} = S_{b}$, respectively.}
\end{center}
\label {fig1}
\end{figure}
We observe that, although the von Neumann entropies $S_{a}$ and $S_{b}$ are always greater than $0$ \textemdash that is, greater than that of the composite state \textemdash over the entire range $0.001 \le c_1 \le 0.999$, as represented by the dotted line in FIG. 1, both the uncertainties 
$\Delta J_{x^\prime}$ and $\Delta J_{y^\prime}$ are greater than $\frac{1}{2}$ in the ranges $0.001 \le c_1 \le 0.258$ and $0.966 \le c_1 \le 0.999$.
The above ranges of $c_1$ correspond to the ranges of $c_2$ as $0.999 \ge c_2 \ge 0.966$, and $0.259 \ge c_2 \ge 0.045$, respectively. Therefore, both the uncertainties of the composite state in Eq. (\ref{3.3}) are greater than those of the individual entangled atoms, which have been represented in Eqs. (\ref{3.46}) to (\ref{3.49}). This is in contradiction with the prevailing idea that the more  the entropy, the greater the uncertainty and disorder. 

Therefore, we see from FIG. 1 that there is violation of the entropic inequalities \cite{Horodecki1}, as in this case we have
\begin{eqnarray}
S_{a,b} < S_a, \label{3.50a1}\nonumber\\
S_{a,b} < S_b, \label{3.50a2}\nonumber
\end{eqnarray}
but
\begin{eqnarray}
\Delta J_{x^\prime_{a,b}} >  \Delta J_{x^\prime_{a}}, \nonumber\label{3.50a3}\\
\Delta J_{x^\prime_{a,b}} >  \Delta J_{x^\prime_{b}}, \nonumber\label{3.50a4}
\end{eqnarray}
and
\begin{eqnarray}
\Delta J_{y^\prime_{a,b}} >  \Delta J_{y^\prime_{a}}, \label{3.50a3}\nonumber\\
\Delta J_{y^\prime_{a,b}} >  \Delta J_{y^\prime_{b}}, \label{3.50a4}\nonumber
\end{eqnarray}
in the ranges
 $$0.001 \le c_1 \le 0.258~~and~~0.966 \le c_1 \le 0.999,$$
$$0.999 \ge c_2 \ge 0.966~~and~~0.259 \ge c_2 \ge 0.045.$$
 We have obtained this feature for the general state of two entangled two-level atoms.

We also observe from FIG. 1  that, when the von Neumann entropies $S_{a} = S_{b}$ of the individual atoms attain their peak values, the uncertainty $\Delta J_{x^\prime}$ of the composite state attains its peak value, while the uncertainty $\Delta J_{y^\prime}$ attains $100 \%$ squeezing. 

\begin{figure}
\begin{center}
\hspace*{-1.0cm}
\includegraphics[width=10cm]{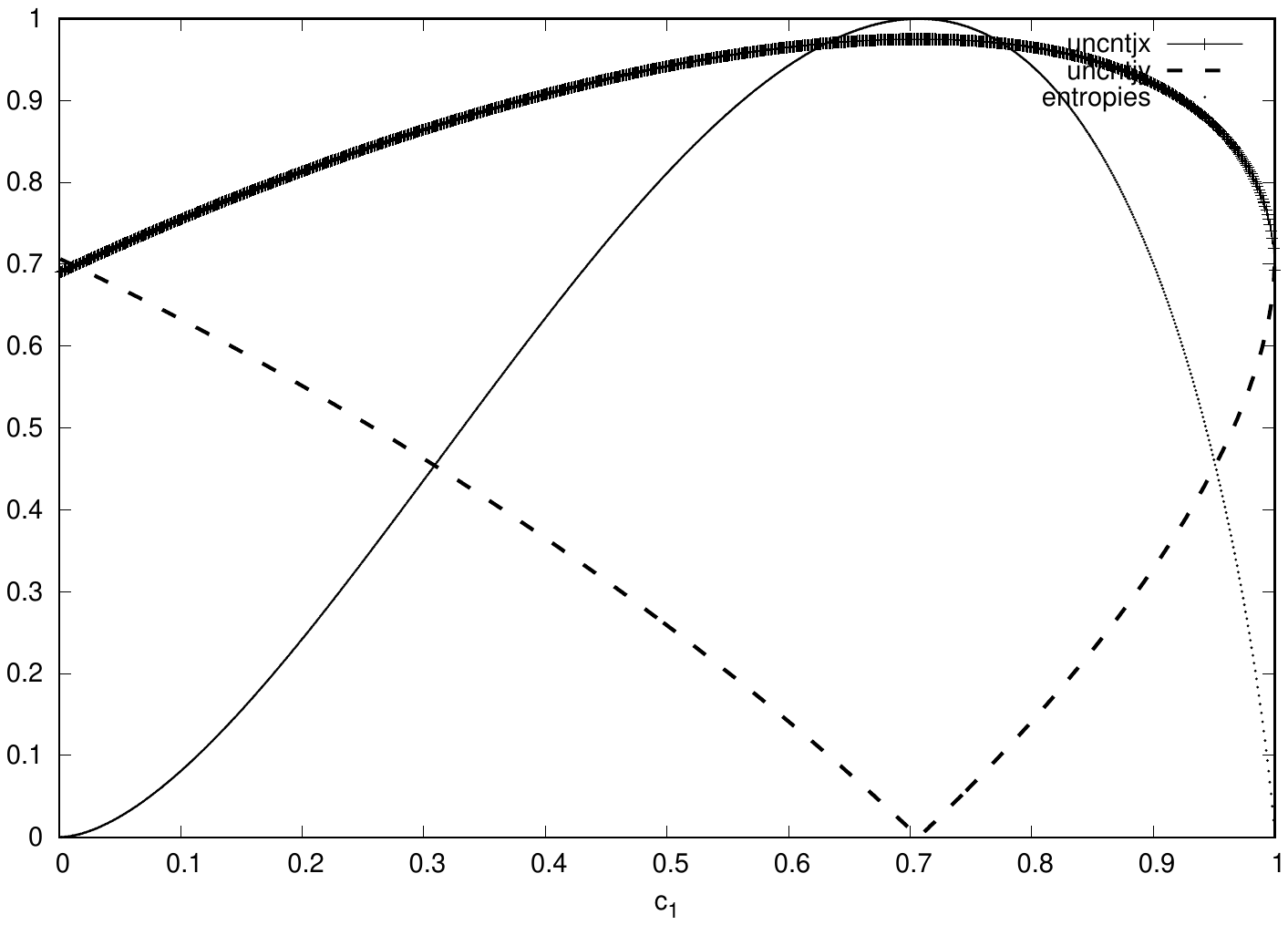}
\caption{Plot of the uncertainties  $\Delta J_{x^\prime}$, $\Delta J_{y^\prime}$, and the von Neumann entropies $S_{a}$ and $S_{b}$ of atoms $a$ and $b$ with respect to the constant $c_1$. Here, we have taken $c_3 = \frac{3}{8}$, $c_4 = \frac{\sqrt{55}}{8}$, $c_5 = \frac{4}{7}$ and 
$c_6 = \frac{\sqrt{33}}{7}$. The constant $c_1$ is plotted along the horizontal axis, while $\Delta J_{x^\prime}$, $\Delta J_{y^\prime}$, and the entropies $S_{a}$ and $S_{b}$ are plotted along the vertical axis. The thick line, dashed line, and  thin line represent $\Delta J_{x^\prime}$, $\Delta J_{y^\prime}$, and $S_{a} = S_{b}$, respectively.}
\end{center}
\label {fig1}
\end{figure}

In FIG. 1, we have chosen the parameters $c_3, c_4, c_5$, and $c_6$ as $\frac{1}{\sqrt{2}}$; that is, the quantum state $|\psi_{ab}\rangle$ is symmetric with respect to these parameters. Now, it can be seen that the above-mentioned feature also holds for asymmetric combinations of these parameters. This is because we observe from Eqs. (\ref{3.29}) and (\ref{3.30}) that while $\Delta J_{x^\prime}$ depends on the parameters $c_1, c_2, c_3, c_4, c_5$, and $c_6$, 
$\Delta J_{y^\prime}$ depends only on $c_1$ and $c_2$. Therefore, variations in the parameters 
$c_3, c_4, c_5$, and $c_6$ do not affect $\Delta J_{y^\prime}$, but they do affect 
$\Delta J_{x^\prime}$. Thus, 
$\Delta J_{y^\prime}$ is always greater than that of the individual entangled atoms in the ranges $0.001 \le c_1 \le 0.258$ and $0.966 \le c_1 \le 0.999$. However, whether $\Delta J_{x^\prime}$ will be greater than that of the individual entangled atoms in the above-mentioned ranges of $c_1$ depends on the choice of the parameters $c_3, c_4, c_5$, and $c_6$. Now, in Eq. (\ref{3.29}), the minimum value of the factor $\sqrt{\Big{(} \frac{1}{2} + c_1c_2 \Big{)}}$ is $0.707$. Therefore, $\Delta J_{x^\prime}$ is greater than $\frac{1}{2}$ \textemdash that is, greater than that of the individual enatngled atoms \textemdash as long as the factor $(c_3c_5 + c_4c_6)$ in Eq. (\ref{3.29}) is greater than $0.707$. Thus, it does not matter whether the parameters $c_3, c_4, c_5$, and $c_6$ are symmetric or asymmetric, as long as their combination makes the factor $(c_3c_5 + c_4c_6)$ in Eq. (\ref{3.29}) greater than $0.707$. As an example, in FIG. 2, we plot $\Delta J_{x^\prime}$, $\Delta J_{y^\prime}$, and the von Neumann entropies $S_{a}$ and $S_{b}$ for an asymmetric combination of the parameters $c_3, c_4, c_5$, and $c_6$, where $c_3 = \frac{3}{8}$, $c_4 = \frac{\sqrt{55}}{8}$, $c_5 = \frac{4}{7}$, and $c_6 = \frac{\sqrt{33}}{7}$. We observe the same result: that both $\Delta J_{x^\prime}$ and $\Delta J_{y^\prime}$, for the composite state, are greater than those of the individual entangled atoms in the same ranges of the parameters $c_1$ and $c_2$, as mentioned previously.

\subsection{IV GENERALIZED RELATION BETWEEN VON NEUMANN ENTROPY AND SQUEEZING PARAMETERS IN TWO-LEVEL ATOMS}
We now establish generalized relationships between the von Neumann entropies of the entangled subsystems of two-level atoms and the spin and spectroscopic squeezing parameters of the composite state. This provides  operational measures for the von Neumann entropies of the entangled subsystems of two-level atoms.

Using Eqs. (\ref{3.2}) and (\ref{3.30}), we obtain
\begin{eqnarray}
c_1^4 - c_1^2 = \Delta J_{y^\prime}^2 - \Delta J_{y^\prime}^4 - \frac{1}{4}.\nonumber
\label{3.51}
\end{eqnarray}
This is a quadratic equation in $c_1^2$, whose roots are
\begin{eqnarray}
c_1^2 = \frac{1}{2} \pm \Delta J_{y^\prime} \sqrt{1 - \Delta J_{y^\prime}^2}.
\label{3.52}
\end{eqnarray}
Therefore, using Eqs. (\ref{3.2}) and (\ref{3.52}), we obtain
\begin{eqnarray}
c_2^2 = \frac{1}{2} \mp \Delta J_{y^\prime} \sqrt{1 - \Delta J_{y^\prime}^2}.
\label{3.53}
\end{eqnarray}
Thus, using Eqs. (\ref{3.50}), (\ref{3.52}), and (\ref{3.53}), we obtain
\begin{eqnarray}
S_{a} &=& S_{b} = -\Big{[}\frac{1}{2} \pm \Delta J_{y^\prime} \sqrt{1 - \Delta J_{y^\prime}^2}\Big{]}
\times\nonumber\\
&&log_2 \Big{[}\frac{1}{2} \pm \Delta J_{y^\prime} \sqrt{1 - \Delta J_{y^\prime}^2}\Big{]} - \nonumber\\
&&\Big{[}\frac{1}{2} \mp \Delta J_{y^\prime} \sqrt{1 - \Delta J_{y^\prime}^2}\Big{]}\times\nonumber\\
&&log_2 \Big{[}\frac{1}{2} \mp \Delta J_{y^\prime} \sqrt{1 - \Delta J_{y^\prime}^2}\Big{]},
\label{3.54}
\end{eqnarray}
or equivalently,
\begin{eqnarray}
S_{a} &=& S_{b} \nonumber\\
&=& -\Big{[}\frac{1}{2} \pm \sqrt{\frac{2}{j}}\Delta J_{y^\prime} \sqrt{\frac{j}{2}} \sqrt{1 - \frac{2}{j}\Delta J_{y^\prime}^2\frac{j}{2}}\Big{]}
\times\nonumber\\
&&log_2 \Big{[}\frac{1}{2} \pm \sqrt{\frac{2}{j}}\Delta J_{y^\prime} \sqrt{\frac{j}{2}} \sqrt{1 - \frac{2}{j}\Delta J_{y^\prime}^2\frac{j}{2}}\Big{]}\nonumber\\
&-&\Big{[}\frac{1}{2} \mp \sqrt{\frac{2}{j}}\Delta J_{y^\prime} \sqrt{\frac{j}{2}} \sqrt{1 - \frac{2}{j}\Delta J_{y^\prime}^2\frac{j}{2}}\Big{]}\times
\nonumber\\
&&log_2 \Big{[}\frac{1}{2} \mp \sqrt{\frac{2}{j}}\Delta J_{y^\prime} \sqrt{\frac{j}{2}} \sqrt{1 - \frac{2}{j}\Delta J_{y^\prime}^2\frac{j}{2}}\Big{]}.\nonumber\\
\label{3.55}
\end{eqnarray}
Here, $j = \frac{N}{2}$, where $N$ is the number of atoms. In this case, we have $N = 2$, and hence 
$j = 1$.

Now, using Eq. (\ref{3.56a1}) in Eq. (\ref{3.55}), and taking $j =1$, we obtain
\begin{eqnarray}
S_{a} &=& S_{b} = -\Big{[}\frac{1}{2} \pm \frac{{\xi_{s_y}}}{\sqrt{2}} \sqrt{1 - \frac{\xi_{s_y}^2}{2}}\Big{]}
\times\nonumber\\
&&log_2 \Big{[}\frac{1}{2} \pm \frac{{\xi_{s_y}}}{\sqrt{2}} \sqrt{1 - \frac{\xi_{s_y}^2}{2}}\Big{]}
\nonumber\\
&-&\Big{[}\frac{1}{2} \mp \frac{{\xi_{s_y}}}{\sqrt{2}} \sqrt{1 - \frac{\xi_{s_y}^2}{2}}\Big{]}\times
\nonumber\\
&&log_2 \Big{[}\frac{1}{2} \mp \frac{{\xi_{s_y}}}{\sqrt{2}} \sqrt{1 - \frac{\xi_{s_y}^2}{2}} \Big{]}.\nonumber\\
\label{3.57}
\end{eqnarray}
This is the generalized relationship between the von Neumann entropies of the entangled subsystems of two-level atoms and the spin squeezing parameter.

To construct a general relation between the von Neumann entropies of the entangled subsystems of two-level atoms and the spectroscopic squeezing parameter, we consider Eq. (\ref{3.58a1}).
In this case, since $N = 2$, we have
\begin{eqnarray}
\xi_{R_y} = \sqrt{2}\frac{\Delta J_{y^\prime}}{|\langle\hat{\mathbf{J}}\rangle|}.
\label{3.59}
\end{eqnarray}
Using Eq. (\ref{3.59}), we rewrite Eq. (\ref{3.54}) as
\begin{eqnarray}
S_{a} &=& S_{b} \nonumber\\
&=& -\Big{[}\frac{1}{2} \pm \xi_{R_y} \frac{|\langle\hat{\mathbf{J}}\rangle|}{\sqrt{2}}\sqrt{1 - \xi_{R_y}^2\frac{|\langle\hat{\mathbf{J}}\rangle|^2}{2}}\Big{]}
\times\nonumber\\
&&log_2 \Big{[}\frac{1}{2} \pm \xi_{R_y} \frac{|\langle\hat{\mathbf{J}}\rangle|}{\sqrt{2}}\sqrt{1 - \xi_{R_y}^2\frac{|\langle\hat{\mathbf{J}}\rangle|^2}{2}}\Big{]}\nonumber\\
&-&\Big{[}\frac{1}{2} \mp \xi_{R_y} \frac{|\langle\hat{\mathbf{J}}\rangle|}{\sqrt{2}}\sqrt{1 - \xi_{R_y}^2\frac{|\langle\hat{\mathbf{J}}\rangle|^2}{2}}\Big{]}\times\nonumber\\
&&log_2 \Big{[}\frac{1}{2} \mp \xi_{R_y} \frac{|\langle\hat{\mathbf{J}}\rangle|}{\sqrt{2}}\sqrt{1 - \xi_{R_y}^2\frac{|\langle\hat{\mathbf{J}}\rangle|^2}{2}}\Big{]}.
\nonumber\\
\label{3.60}
\end{eqnarray}
Thus, we have established a generalized relationship between the von Neumann entropies of the entangled subsystems of two-level atoms and the spectroscopic squeezing parameter.
The Eqs. (\ref{3.57}) and (\ref{3.60}) provide generalized operational measures of the von Neumann entropies of the entangled atoms.

In FIG. 3, we plot the spin and spectroscopic squeezing parameters of the composite state, and the von Neumann entropies of the entangled subsystems of two-level atoms, with respect to the constant 
$c_1$.

\begin{figure}
\begin{center}
\hspace*{-1.0cm}
\includegraphics[width=9cm]{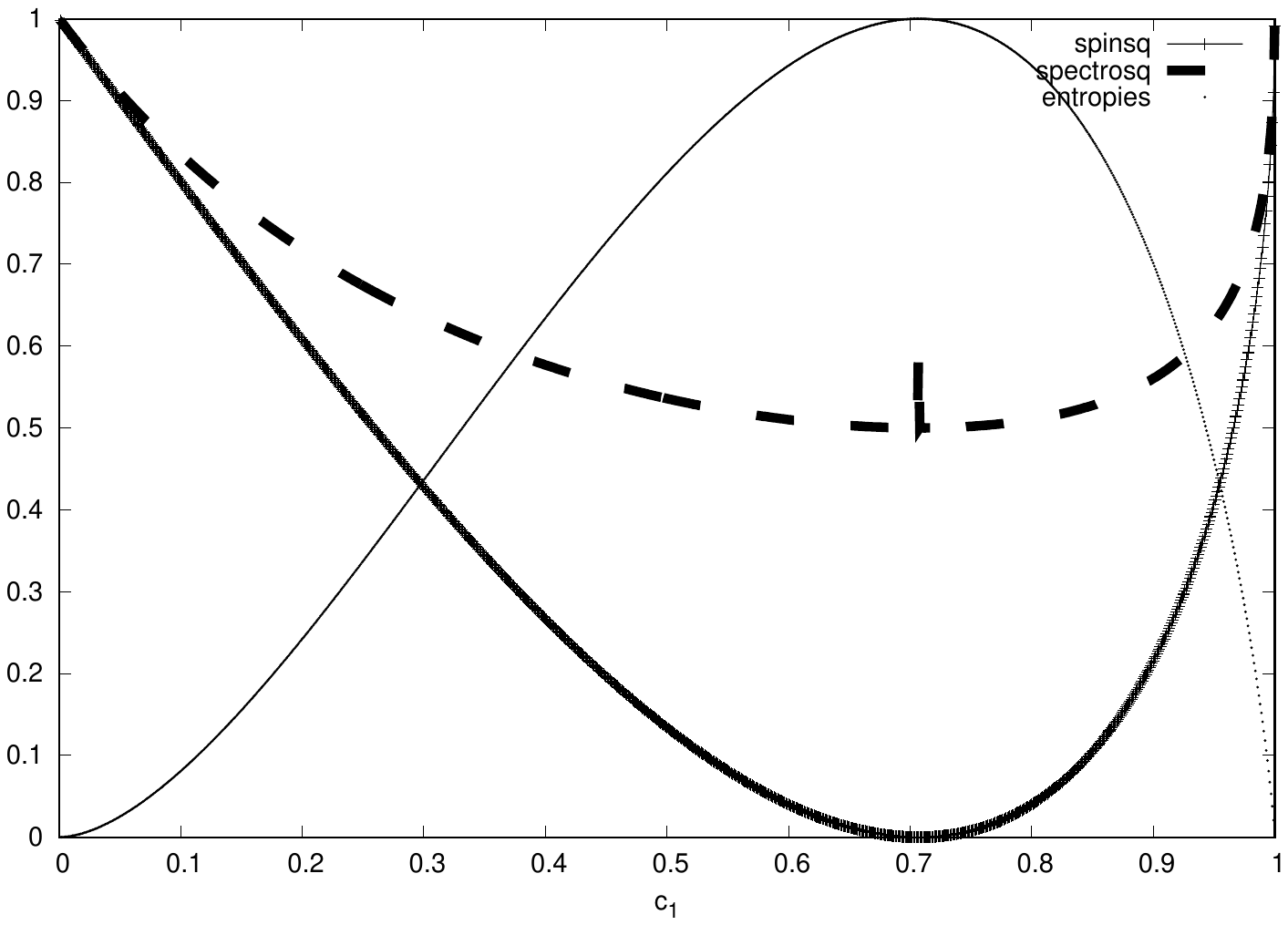}
\caption{Plot of the spin squeezing parameter $\xi_{s_y}^2$, spectroscopic squeezing parameter 
$\xi_{R_y}^2$, and the von Neumann entropies $S_{a}$ and $S_{b}$ of the entangled parts  with respect to the constant $c_1$. Here, we have taken $c_3 = c_4 = c_5 = c_6 = \frac{1}{\sqrt{2}}$. The constant $c_1$ is plotted along the horizontal axis, while $\xi_{s_y}^2$, $\xi_{R_y}^2$, and $S_{a} = S_{b}$ are plotted along the vertical axis. The thick line, dashed line, and dotted line represent 
$\xi_{s_y}^2$, $\xi_{R_y}^2$, and $S_{a} = S_{b}$, respectively. }
\end{center}
\label {fig3}
\end{figure}
We see from FIG. 3 that as the spin squeezing and spectroscopic squeezing of the composite state increase, the von Neumann entropies of the entangled parts also increase, and when there is $100\%$ spin squeezing of the composite state, the von Neumann entropies of the entangled parts attain their peak values. Therefore, the greater the squeezing of the composite state, the larger the von Neumann entropies of the entangled parts. Then, as the squeezing of the composite state decreases, the von Neumann entropies of the entangled parts also decrease. This feature, as we can see, applies to the general state of two two-level atoms given in Eq. (\ref{3.3}).
\subsection{V CONCLUSION}
We have analytically calculated  the uncertainties for a general state of two entangled two-level atoms, both for the composite state and the entangled parts. We observe that the von Neumann entropies of the entangled parts are always greater than that of the composite state. However, for certain ranges of the superposition constants of the quantum state, the uncertainties of the entangled parts are found to be less than those of the composite state. This observation does not align with the prevailing idea that greater entropy implies greater uncertainty. As we can see, this feature holds for the general state of two two-level atoms. 

We have also analytically established generalized relationships between the von Neumann entropies of the entangled parts of two-level atoms and the uncertainties, spin squeezing, and spectroscopic squeezing parameters 
\textemdash used in the context of Ramsey spectroscopy \textemdash of the composite state. This provides  generalized operational measures of the von Neumann entropies of the entangled atoms.

\end{document}